\documentclass[a4paper]{jpconf}
\usepackage{iopams}

\usepackage{cases}
\usepackage{makeidx}
\makeindex
\usepackage {graphicx}
\usepackage{color}

\def\>{\right\rangle}
\def\<{\left\langle}
\def\be{\begin{equation}}
\def\ee{\end{equation}}
\def\ba{\begin{array}{lll}}
\def\ea{\end{array}}

\def\beq{\begin{eqnarray}}
\def\eeq{\end{eqnarray}}

\begin{document}
\title{Finite frequency noise for Laughlin state investigated by a resonant circuit}
\author{M Carrega$^1$, D Ferraro$^{2,3}$, A Braggio$^1$ and M. Sassetti$^{4,1}$}
\address{$^1$ SPIN-CNR, Via Dodecaneso 33, 16146 Genova, Italy}
\address{$^2$ Aix Marseille Universit\'e, CNRS, CPT, UMR 7332, 13288 Marseille, France}
\address{$^3$ Universit\'e de Toulon, CNRS, CPT, UMR 7332, 83957 La Garde, France}
\address{$^4$ Dipartimento di Fisica, Universit\`a di Genova, Via Dodecaneso 33, 16146 Genova, Italy}
\ead{alessandro.braggio@spin.cnr.it}

\begin{abstract}
We study the finite frequency (F.F.) noise properties of edge 
states in the Laughlin state. We investigate the model of a resonant detector coupled to a 
quantum point contact in the weak-backscattering limit.
In particular we discuss the impact of possible renormalization of the Luttinger exponent, due to environmental effects, on the measured quantities and we propose a simple way to extract such non-universal parameters from noise measurements.
\end{abstract}

\section{Introduction}
Strong correlations in low dimensional systems can lead to very intriguing effects such as charge fractionalization and non-Fermi liquid behaviour.
A remarkable example is the fractional quantum Hall effect~\cite{Dassarma97}, where  quasiparticles carry fractional charges and have non-trivial statistics~\cite{Wen95, Kane94, Kane95, Laughlin83, Jain89}.
Transport measurements in a quantum point contact (QPC) geometry are among the simplest ways to probe these peculiar properties.
In particular, by studying current fluctuations in the shot noise limit, one can directly measure the charge of the excitations involved in the tunnelling through the QPC~\cite{Depicciotto97, Saminadayar97, Reznikov99}.
Indeed, the zero frequency current-current correlation in the weak-backscattering regime, is predicted to be proportional to the induced backscattering current via the fractional charge associated to the tunnelling excitation between the opposite edges of the Hall bar~\cite{Wen95,Kane94,Kane95, Martin05}. Clear experimental signatures of this fact have been reported for the Laughlin sequence~\cite{Laughlin83}, with filling factor $\nu=1/(2n+1)$ ($n \in \mathbb{N}$). Here the measured fractional charge is $e^{*}= e/ (2n+1)$ ($e$ the electron charge), in agreement with theoretical predictions~\cite{Depicciotto97, Saminadayar97,Reznikov99}.\\
In the case of composite edges, such as in the Jain sequence~\cite{Jain89} with $\nu = p/(2 n p + 1)$ ($p \in \mathbb{Z}$), the situation is more involved and, at low energies, various excitations with different fractional charges can contribute to the tunnelling.
Indeed, different experiments have reported the observation of a crossover in the value of the effective charge (as a function of temperature) which has been explained in terms of competition between different excitations, namely the single quasiparticle (QP) (with fundamental charge $e^* = \nu e/|p|$) and $p$- agglomerates (with charge $\nu e$)~\cite{Chung03, Bid09, Dolev10, Ferraro08, Ferraro10, Ferraro10b, Carrega11}.
Furthermore, renormalizations of the chiral Luttinger liquid exponents are usually needed to fully explain this phenomenology~\cite{Chung03, Bid09, Dolev10, Ferraro08, Ferraro10, Ferraro10b, Carrega11, Braggio12}, and many mechanisms responsible for these renormalizations were proposed in literature~\cite{Braggio12, Rosenow02, Papa04, Mandal02, Yang03}.
Even for the simple Laughlin sequence renormalization parameter for the Luttinger exponent are often introduced to fully explain anomalous current-voltage characteristics reported at very low temperature~\cite{Roddaro03, Roddaro04}.\\
An important tool in order to gain more information about the properties of fractional excitations is represented by finite frequency current correlations~\cite{Blanter00, Rogovin74, Chamon95}.
For example, for quantum Hall QPC transport, the F.F. noise is predicted to show resonances in correspondence of Josephson frequencies, which are proportional to the fractional charges~\cite{Rogovin74,Chamon95, Chamon96, Dolcini05, Bena07, Carrega12, Ferraro12}.
In this case the presence of fractionally charged tunnelling excitations may be revealed, at extremely low temperatures, by the presence of peaks or dips in the noise spectrum occurring at frequency $\omega\backsim  e^{*} V/\hbar$.
The study of F.F current fluctuations~\cite{Carrega12, Ferraro12, Ferraro14} can also serve as an alternative tool to probe the power-law behaviour expected for these chiral Luttinger liquids.\\
However, at these energy scales the detection scheme has to be considered with care, in order to properly identify which quantity is effectively probed in real experiments.\\ 
In this context Lesovik and Loosen~\cite{Lesovik97} introduced a model based on a resonant LC circuit as prototypical scheme for F.F. noise measurement.
It has been shown that the measured quantity for the LC detector setup can be expressed in terms of the non-symmetrized F.F. noise which reflects the emission and adsorption contributions of the active system under investigation, {\it i.e.} the QPC~\cite{Ferraro14, Lesovik97, Gavish00, Bednorz13}.
The non-symmetrized noise has been considered in literature for different systems as the ultimate source of information of quantum noise properties~\cite{Aguado00, Sukhorukov01, Johansson02, Deblock03, Billangeon06, Safi08}.\\
Here we consider the F.F. detector output power of a resonant circuit coupled to a QPC in the fractional quantum Hall regime, as discussed in Ref.~\cite{Ferraro14}. We analize symmetric and  measured noise as a function of bias and temperature at a fixed frequency $\omega$. For sake of simplicity we limit the discussion to the case of the Laughlin sequence with $\nu=1/(2n+1)$.
In particular we will consider possible renormalizations of the Luttinger exponent, due to the interaction with the external environment, extending the previous work of Ref.~\cite{Ferraro14}, to investigate the robustness of the relevant features associated to F.F. current fluctuations and clarify the range of validity of the results discussed in Ref. \cite{Ferraro14}.

\section{Theoretical model}

The field theoretical description of the edge states in the Laughlin sequence~\cite{Laughlin83}, with filling factor $\nu=1/(2n+1)$ is given in terms of a single bosonic mode through the Lagrangian density 
\be
\mathcal{L}=-\frac{1}{4 \pi \nu}  \partial_{x}\varphi  ( \partial_{t} +v_{c} \partial_{x} ) \varphi,
\label{free_Laughlin}
\ee
where $\varphi(x)$ is a right-moving field with commutation relation 
\be
\label{commutazione_lau}
\left[\varphi(x) , \varphi(x') \right] = i \pi \nu \mathrm{sign}(x-x')~.
\ee
This bosonic mode propagates along the edge at velocity $v_{c}$ and is related to the electron particle density by
\be
\rho(x)=\frac{1}{2 \pi} \partial_{x} \varphi (x).
\label{density}
\ee
It is possible to show that the single QP is the most relevant excitation in the tunnelling process \cite{Kane95}  and, using bosonization techniques \cite{Miranda03}, we can express the single QP operator as
\be 
\Psi(x)= \frac{\mathcal{F}}{\sqrt{2\pi \alpha}} e^{i  \nu \varphi (x)}
\ee
being $\alpha$ a finite length cut-off and $\mathcal{F}$ the so-called Klein factor~\cite{Martin05}.
The fundamental charge associated to this excitation is $ e^{*}=\nu e$, with scaling dimension~\cite{Kane95}

\be
\Delta= g \frac{\nu}{2}   
\label{Delta_lau}
\ee

being $g\geq 1$ a possible renormalization factor associated to external interactions like, among the others, the coupling with one dimensional phonon mode, the Coulomb interaction at the QPC or the interplay between $1/f$ noise and dissipation generated by the circuit~\cite{Braggio12, Rosenow02, Papa04, Mandal02, Yang03}. 

\subsection{Measured finite frequency noise}
\begin{figure}[ht]
\centering
\includegraphics[scale=.50]{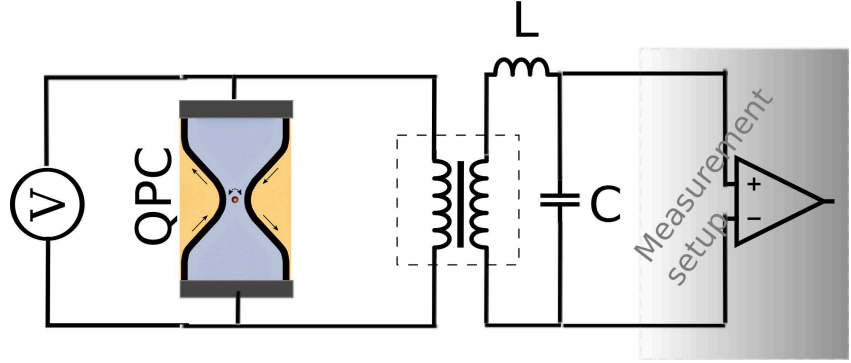}
\caption{Schematic view of an Hall bar (blue) with a QPC coupled with LC detector. A bias voltage $V$ is applied to the QPC and QP excitations can tunnel between the edges. The two circuits are impedance matched via a coupling circuit (inside dashed line).}
\label{fig:1}
\end{figure}
We consider  the F.F. backscattering current fluctuations in a single 
QPC geometry coupled to the detector setup shown in Fig.~\ref{fig:1}.
Here the QPC is subjected to a bias voltage V and
coupled to a resonant LC circuit, the detector (with frequency $\omega=\sqrt{1/LC}$), via an impedance matching circuit 
 (inside the dashed line in the figure)~\cite{Altimiras14,Altimiras14a}.
We assume a small electrical coupling $K \ll 1$ between the QPC and the detector~\cite{Ferraro14, Lesovik97}, with a very high quality factor and we keep constant the resonant frequency $\omega$ as usually done in experiments. The point-like tunnelling process of single QP between the upper and lower edge of the Hall bar  through the QPC can be described by the tunnelling Hamiltonian
\be
\label{tun}
\hat{{\cal H}}_{T}= t\  \Psi(0){\Psi^{\dagger}}(0)+h.c.\ ~,
\ee
where $t$ represents the tunnelling amplitude, which is assumed energy-independent for sake of simplicity.
A finite bias voltage $V$ is included in the tunnelling amplitude through the usual gauge transformation $t \to  te^{i\omega_0t}$, where $\omega_0=e^*V/\hbar$ is the Josephson resonance 
associated to the fundamental charge $e^*$~\cite{Martin05}.\\
The non-symmetrized noise~\cite{Lesovik97, Gavish00, Aguado00} is defined in terms of the backscattering current fluctuations
\be
S_{+}(\omega)=\frac{1}{2}\int^{+\infty}_{-\infty} dt \, e^{i \omega t}
 \langle \delta I_{B}(0) \delta I_{B}(t) \rangle, 
 \label{S_plus}
\ee
and represents, for $\omega>0$, the noise emission of the system into the detector. 
In the above expression we considered the back-scattering current fluctuation $\delta I_B= I_{B}-\langle I_{B}\rangle$ with the average $\langle...\rangle$ taken over the quantum statistical ensemble. Using the time-reversal symmetry properties of Eq.~(\ref{S_plus}) we can deduce the absorptive part of the spectrum as
\be
\label{TR}
S_{-}(\omega)=S_{+}(-\omega)\ .
\label{Plus_minus}
\ee 
Usually the theoretical study of F.F. properties relies on the so-called symmetrized noise, {\it i.e.}
\be
S_{sym}(\omega) = S_{+}(\omega)+S_{-}(\omega)~,
\label{eq:s_sym}
\ee
however the measurable quantity in the scheme of Fig.~\ref{fig:1} is the output power proportional to the variation of the energy stored in the LC before and after the switching on of the LC-QPC coupling.
In the following we refer to it as measured noise $S_{{meas}}(\omega)$ and, in terms of the non-symmetrized noise, it reads
\be
S_{meas}(\omega)= K \left\{S_{+}(\omega)+n_{B}(\omega) \left[ S_{+}(\omega)-S_{-}(\omega)\right]\right\}\ ,
\label{S_meas}
\ee
with $n_{{\rm B}}(\omega ) = \left[e^{\beta_{{ c}}\omega}-1\right]^{-1}$ the boson distribution and $\beta_{c}=1/k_{B}T_{c}$ the detector inverse temperature.
We recall that, at the lowest order in the tunnelling Hamiltonian, the non-symmetrized emission noise can be written in terms of the single QP tunnelling rate as~\cite{Martin05, Gavish00}
\be
S_{+}(\omega, \omega_{0})= \frac{( e^{*})^{2}}{2} \left[{\bf{\Gamma}}\left(-\omega+ \omega_{0}\right)+{\bf{\Gamma}}\left(-\omega- \omega_{0}\right)\right],
\label{S_plus_rate}
\ee
where
 \be
 \label{rate}
 {\bf{\Gamma}}(E) = |t|^{2} \int^{+\infty}_{-\infty} d \tau\  e^{i E t} \mathcal{G}^{<}_{ -}(-t) \mathcal{G}^{>}_{ +}(t),
 \ee
with $\mathcal{G}^{>}_{ \pm}(t)=\langle\Psi_{\nu,\pm}(t)\Psi_{\nu,\pm}^{\dagger}(0) \rangle=(\mathcal{G}^{<}_{ \pm}(-t))^*$ the greater/lesser correlation functions of the single QP  for the edge $j= \pm$.
For the Laughlin sequence these correlation functions are~\cite{Ferraro08, Ferraro10}
 \be
 \label{Ggreater}
\mathcal{G}^{>}_{\pm}(t) =\frac{1}{2\pi \alpha} \left[\frac{|\Gamma(1+(\beta \omega_{c})^{-1}-i t /\beta )|^{2}}{\Gamma^{2}( 1+(\beta \omega_{c})^{-1}) (1\pm i \omega_{c} t)} \right]^{2\Delta}\!\!\!\!\!\!\!\!,  
\ee
being $\Gamma(x)$ the  Euler gamma function and $\omega_{c}=v_{c}/a$ the high frequency cut-off. This leads, in the zero temperature limit, to the simple expression \cite{Guinea95, Cuniberti96}
\be
{\bf{\Gamma}}(E) =\frac{|t|^2}{2\pi \alpha^{2}} \left(\frac{E}{\omega_{c}} \right)^{4 \Delta} \frac{E^{-1}}{\Gamma(2 g \nu)} \Theta(E)\propto \Theta(E) E^{4\Delta-1}
\label{Rate_zeroT}
\ee
with $\Theta(x)$ the Heaviside step function. 

It is worth to note that the exponent depends on the scaling dimension of the QP operator and hence is affected by the renormalization parameter $g \geq 1$.

\section{Results and discussion}
In the Ref.~\cite{Ferraro14} one can find a complete discussion of the proposed detection scheme in the general contest of the fractional quantum Hall states. Here we limit only to discuss F.F. symmetrized and measured noise in the Laughlin sequence. For sake of clarity we focus on the state at  $\nu=1/3$ which, as stated before, carries a tunnelling charge $e^* = e/3$. For this state we investigate the effects of the renormalization of the Luttinger parameter on the behaviour of the physical quantities. We consider the very low temperature regime for both the QPC and the detector ($T= T_{c}\sim 15$~mK)\footnote{Note that for $T=T_c$ the measured noise coincides with the excess noise power $S_{ex}(\omega , \omega_0 ) = S_{meas}(\omega , \omega_0 ) - S_{meas}(\omega , \omega_0 =0)$, a quantity that is extracted keeping fixed the coupling $K$. This is experimentally easier since the critical step is the frequency dependence of the fine tuning in the impedance matching circuit.} and we keep fixed the frequency $\omega$ scanning on the bias $\omega_0$. This will allow us to be closer to realistic experimental conditions.\\
\begin{figure}[ht]
\centering
\includegraphics[scale=.45]{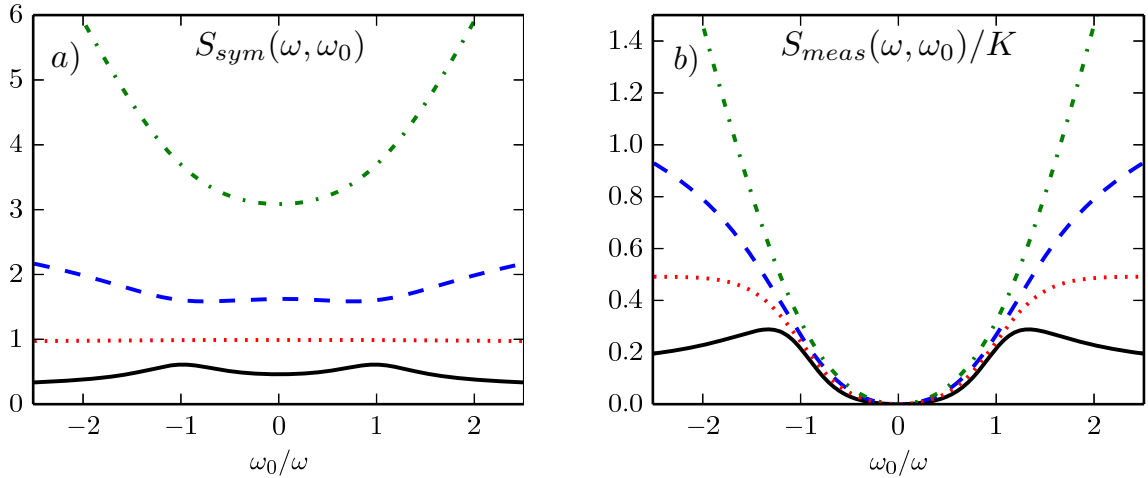}
\caption{(Color online) a) Symmetrized noise $S_{sym}(\omega, \omega_{0})$ (in units of $S_{0}= e^{2} |t|^{2}/((2\pi \alpha)^2 \omega_{c}) $) as a function of $\omega_{0}/\omega$ and b) measured noise $S_{meas}(\omega, \omega_{0})/K$ (in units of $S_{0}$) as a function of $\omega_{0}/\omega$. The figures are for $\nu=1/3$ and for different renormalizations of the Luttinger parameter: $g=1$ black full, $g=1.5$ red dotted, $g=2$ blue dashed and $g=3$ green dashed dotted. Other parameters are: $T=T_{c}=15$~mK, $\omega = 60$~mK.} 
\label{fig:2}
\end{figure}
In Fig. \ref{fig:2} a) we show the behaviour of $S_{sym}$ as a function of $\omega_{0}/\omega$. In absence of renormalization ($g=1$, black full curve) we observe a two peaks structure associated to the resonances at the Josephson frequency ($\omega_{0}/\omega=\pm 1$) \cite{Carrega12, Ferraro14}. By increasing the renormalization we first find a completely flat regime ($g=1.5$, red dotted curve) that turns into a double-dipped structure ($g=2$, blue dashed curve). At even higher renormalization ($g=3$, green dashed dotted curve) every feature associated to the Josephson resonances is washed out and the signal increase at high voltage with a progressively increasing power-law. For not so strong renormalization this phenomenology can be easily explained by looking at the behaviour of the noise near the resonances. Indeed, according to Eq. (\ref{Rate_zeroT}), around these peaks ($\omega_{0}/\omega\approx \pm 1$), up to a smearing of the singularities associated to thermal effects, the noise is symmetric and scale as 
\be
S_{sym}(\omega, \omega_{0})\approx |\omega_{0}-\omega|^{2 \nu g-1}
\ee 
in agreement with what shown in Fig. \ref{fig:2} a). 
However, when the power-law growth is too strong, all the relevant features near Josephson resonances are washed away. 
 
Another remarkable characteristic of $S_{sym}$ is its finite value at $\omega_{0}/\omega\approx 0$. This contribution, which scale as a power law with temperature, is essentially due to the so-called zero point fluctuations \cite{Lesovik97} which are taken into account by the symmetric combination of the tunnelling rates in Eq. (\ref{eq:s_sym}). 

Different is the situation for what it concerns $S_{meas}$ shown in Fig. \ref{fig:2} b). In this case all the curves start from zero at zero voltage due to the absence of contribution from zero point fluctuations. All curves are very similar for $|\omega_0/\omega | \leq 1$, where the physics is essentially dominated by thermal effects. Then, for $\omega_0 -\omega  \gg k_B T$, they follow a similar power-law behaviour around the resonance   
\be
S_{meas}(\omega, \omega_{0})\approx (\omega_{0}-\omega )^{2 \nu g-1}.
\ee
as already discussed for the symmetrized noise. 
 
\begin{figure}[ht]
\centering
\includegraphics[scale=.40]{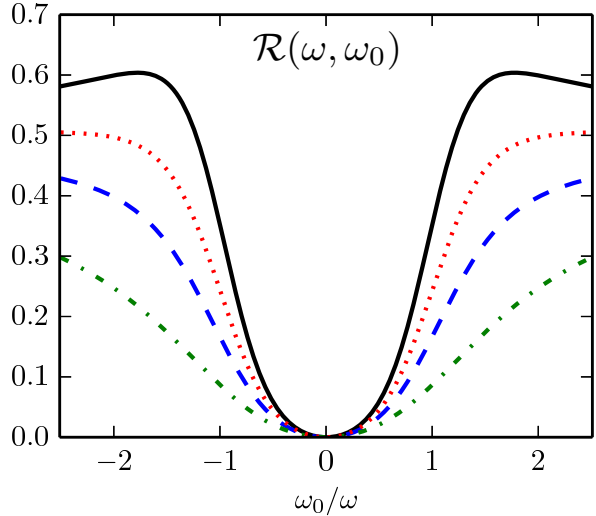}
\caption{(Color online) Ratio $\mathcal{R}(\omega, \omega_{0})$ between $S_{meas}(\omega, \omega_{0})/K$ and $S_{sym}(\omega, \omega_{0})$ for $\nu=1/3$ and different values of the renormalization: $g=1$ black full, $g=1.5$ red dotted, $g=2$ blue dashed and $g=3$ green dashed dotted. Other parameters are: $T=T_{c}=15$ mK.} 
\label{fig:3}
\end{figure}

In order to better understand the main differences between $S_{meas}$ and $S_{sym}$ it is useful to consider the ratio 
\be
\mathcal{R}(\omega, \omega_{0})= \frac{S_{meas}(\omega, \omega_{0})}{K S_{sym}(\omega, \omega_{0})}.
\ee
In the zero temperature limit ($T=T_{c}\to0$) this quantity reduces to  

\be
\mathcal{R}(\omega, \omega_{0})\approx \frac{S_{+}(\omega, \omega_{0})}{\left[S_{+}(\omega, \omega_{0})+S_{-}(\omega, \omega_{0})\right]}.
\label{Ratio}
\ee
Using the expression for the tunnelling rate at zero temperature in Eq. (\ref{Rate_zeroT}), one obtains 
\beq
\mathcal{R}(\omega, \omega_{0})\approx \left\{
\begin{array}{cc}
0 & |\omega_{0}/\omega|<1\\
\left[\frac{1}{2}+\left(\frac{1}{2}-g \nu\right)|\frac{\omega}{\omega_{0}}|\right] & |\omega_{0}/\omega|\gg 1\\
\end{array}
\right..
\eeq
At finite temperature, keeping $T=T_{c}$ these two asymptotic behaviours remains valid respectively for $\omega_{0}/\omega \approx 0$ and $\omega_{0}/\omega \gg 1$ as clearly shown is Fig. (\ref{fig:3}), while the other regions of the curves are rounded due to thermal effects. 
According to the previous considerations, at extremely high voltage with respect to the measurement frequency the ratio reaches the asymptotic value $\mathcal{R}\approx 1/2$ and the way to approach it, namely from above or from below, crucially depends on the renormalization parameter. In particular for $\nu<1/2g$  (black full curve) this ratio presents two maxima symmetric with respect to $\omega_{0}/\omega=0$, while for $\nu>1/2g$ the curves increase (decrease) monotonically for positive (negative) voltage. Expression in Eq. (\ref{Ratio}) therefore suggests that measurement of $S_{sym}$ and $S_{meas}$ in the same setup could provide an estimation of the non universal renormalization of the Luttinger parameters.\footnote{Note that, in general, the measurement of $S_{sym}$ requires a different detection scheme.\cite{Blanter00,Dolcini05,Bednorz13}}
\section{Conclusions}
We have studied the finite frequency (F.F.) noise properties of edge states in the Laughlin state. We have considered a realistic measuring setup based on a resonant circuit coupled to a quantum point contact in the fractional Hall regime.
Symmetric and measured F.F. noise have been investigated,  properly discussing the effects of possible renormalization of the Luttinger exponents on these observables, mainly focusing on the robustness of the relevant peaked or dipped structures. We finally considere their ratio as a useful way to access the value of the non universal renormalization parameter.
\section*{Acknowledgements}
 We acknowledge the support of the MIUR-FIRB2012 - Project HybridNanoDev (Grant  No.RBFR1236VV) and the EU FP7/2007-2013 under REA
grant agreement no 630925 -- COHEAT. 

\end{document}